\documentclass[epj]{svjour}
\usepackage{braket}
\usepackage{amsmath}
\usepackage{graphicx}
\usepackage{color}

\begin{document}

\title{The isotropic Compton profile difference across the phase transition of VO$_2$ }

\author{Kari O.\, Ruotsalainen \inst{1}, Juho Inkinen \inst{1}, Tuomas Pylkk{\"a}nen \inst{1}, Thomas Buslaps \inst{2}, Mikko Hakala \inst{1}, Keijo H{\"a}m{\"a}l{\"a}inen \inst{1}
\thanks{\emph{Present address:} University of Jyv{\"a}skyl{\"a}, Finland}, 
 and Simo Huotari \inst{1}
}
\institute{Department of Physics, P. O. Box 64, FI-00014 University of Helsinki, Finland \and ESRF --- The European Synchrotron, 71 avenue des Martyrs, CS40220, 38043 Grenoble Cedex 9}
\date{Received: date / Revised version: date}

\abstract{
We studied the isotropic Compton profile of the prototypical oxide VO$_2$ 
across the temperature induced electronic and structural phase transition 
at T$_\textrm{C}\approx$ 340 K. We show that the phase transition leaves an 
observable signal, which facilitates Compton scattering 
studies of electronic structure and 
phase transitions in complex solids in powder form.  
We compare the experimental 
results with density functional theory calculations and find agreement 
in the shape of the difference profile, although 
the amplitude of the observed features is overestimated. 
The origin of the disagreement is discussed and we argue that it 
mainly originates mostly correlation effects
beyond our current calculations and possibly 
to some extent, from thermal motion.
\PACS{
      {71.30.+h}{Metal-insulator transitions and other electronic transitions}   \and
      {71.18.+y}{Fermi surface: calculations and measurements}
     } 
} 
\authorrunning{K.\, O.\, Ruotsalainen et al.}
\titlerunning{The isotropic Compton profile differences across the phase transition of VO$_2$}
\maketitle

\section{Introduction}

Inelastic X-ray scattering at high energy and momentum 
transfers, Compton scattering, probes the electron momentum density $\rho(\vec{p})$ (EMD). \cite{eisenberger70}
Compton scattering has been shown to be an
advantageous tool in obtaining bulk sensitive quantitative information 
on electronic structure. \cite{barbillieni09,sakurai11,hiraoka05,hiraoka07,shukla99,huotari2010}
In the framework of the impulse approximation,\cite{eisenberger70,eisenberger74} 
the Compton scattering cross section
is proportional to the EMD projected onto the scattering vector $\vec{q}$.
 This quantity is known as the Compton profile (CP), J$(\vec{p_z})$. 
The EMD is a ground state property and thus useful for comparisons
with theoretical calculations as the computation of response functions, 
with the associated approximations, is avoided. 

Metal-insulator transitions (MIT) present a rich topic of study as they are observed 
in a wide range of compounds and often result in surprising discoveries. \cite{imada98} 
This is especially true in the case of strongly correlated electron systems,
 where in addition to the canonical Mott transition, phase transitions induced by e.g. 
temperature, pressure or doping lead into novel electronic ground states, such as high 
temperature superconductivity in the cuprates or colossal magnetoresistance in the 
manganites. MIT exhibiting materials are often complex oxides, in which the 
interpretation of various experiments can be challenging. It it thus important to apply 
multiple complementary probes to a given problem. VO$_2$ exhibits a complex MIT where 
the electronic phase transition takes place at T$_{\textrm{C}}$ $\approx$ 340 K and coincides with a 
structural transition from a tetragonal metallic R phase above T$_{\textrm{C}}$ to a monoclinic insulating M1 phase
in pristine samples. \cite{morin1959,marezio72,blaauw75}  
The respective roles of electron correlations and electron phonon coupling effects 
in inducing  the MIT have been studied with numerous experimental and theoretical 
approaches. 
\cite{blaauw75,shin90,suga09,eguchi08,haverkort05,biermann05,wentzcovitch94,kim13,eyert11,continenza99,gatti07,grau-crespo12} 

From the experimental perspective, VO$_2$ is an attractive prototype material 
for investigating phase transitions with Compton scattering. This is due to the fact
that in comparison with e.g. MIT-exhibiting ternary oxides, 
VO$_2$ has a larger valence-to-core electron-number ratio.
Compton scattering studies of EMD anisotropies 
across MIT boundaries have been presented for La$_{2 - 2x}$Sr$_{1 + 2x}$Mn$_2$O$_7$, La$_{2 − x}$Sr$_x$CuO$_4$, 
Ba$_{1 - x}$K$_x$BiO$_3$ and PrBa$_2$Cu$_3$O$_{7 - \delta}$. \cite{barbillieni09,sakurai11,hiraoka05,hiraoka07,shukla99} 
The temperature dependence of Compton profiles in weakly correlated systems 
has also been studied.\cite{dugdale98,sternemann01,huotari02,erba15} Furthermore, Compton profiles across solid-liquid phase boundaries in B and Si 
have been studied 
with Compton  scattering and ab-initio molecular dynamics. \cite{okada12,okada15}
The effects of temperature and the metal-superconductor 
phase transition in the Compton profile of MgB$_2$ 
have been studied experimentally and computationally. \cite{nygard04,joshi07}

The developments presented in the above-mentioned studies suggest that 
temperature-dependent in situ Compton scattering studies of MITs may now be initiated.
Our contribution to this promising line of inquiry is a study on the 
isotropic EMD difference across the MIT phase boundary in VO$_2$. 
In particular, we demonstrate the ability of Compton scattering to reveal a MIT-signal in
a powder averaged experiment. We compare the experimental result to density functional theory 
calculations using local exchange-correlation approximations to yield a 
first order approximation for the difference Compton profile,
as it has been shown that the local density approximations yield
reasonable anisotropy Compton profiles in e.g. Ni metal. \cite{chioncel14}  
We compare the experimental Compton profile difference 
across the MIT boundary to density functional theory calculations. Good agreement
is found on the shape of the difference profile, but the amplitude is overestimated. 
To analyze the origin of the discrepancy we compare the relative orders of 
magnitudes of physical effects not included in the calculations.

\section{Experiment}

\begin{figure}
\includegraphics[width=\linewidth]{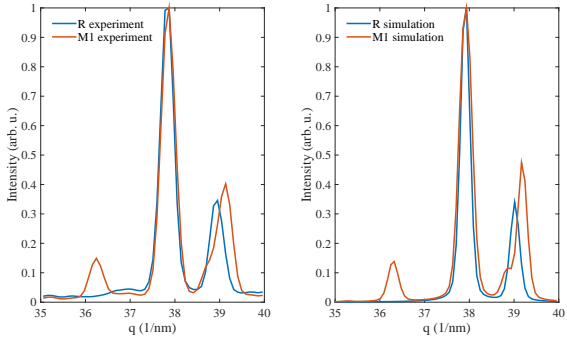}
\caption{\label{fig:xrd} Portions of the experimental (left) and simulated (right) X-ray diffraction patterns
for the insulating and metallic structures VO$_2$. We present a momentum transfer region in which there are
reflections characteristic to the M1 phase, which clearly demonstrate the occurrence of the phase transition. 
}
\end{figure}

The Compton scattering experiment was performed at 
the beamline ID15B of the ESRF. The incident beam 
was monochromatized to a photon energy of 87 keV with 0.1\% bandwidth 
using a bent Si(511) monochromator in the Laue geometry. 
A 12-element Ge solid state detector at a nominal scattering angle of 154$^\circ$ was used 
to record the spectra. The total momentum resolution 
of the experiment was 0.7 atomic units (a.u. from hereon), 
and was calculated from the detector resolution, incident bandwidth and 
scattering solid angles. We used a N$_2$ gas blower as the sample heater.
We monitored the stability of the temperature with a thermocouple placed approximately 1 cm from the sample
and  found that it was  stable within $\pm$ 2 K. The temperatures discussed from hereon refer to
the heater setting. 
A pressed 2 mm thick pellet of commercial 99\% grade VO$_2$ powder 
(Alfa Aesar) was used as the sample. We first performed temperature dependent 
x-ray diffraction measurements and observed that the structural phase transition 
was initiated at a heater setting of 377 K and the sample had fully changed
phase by 384 K. We thus chose the Compton profile measurement temperature of the M1
phase at 363 K. The Compton profile of the R phase was measured at 383K.  
We also performed further x-ray diffraction measurements while recording
the Compton profiles. The diffraction patterns agreed with the aforementioned 
initial test measurements at and were stable throughout the Compton profile measurements. 

In Fig. \ref{fig:xrd} 
we present portions of experimental diffraction patterns taken during the
Compton profile measurements and 
a comparison with calculated powder patterns using the structures 
from McWhan et al (R) and Longo and Kierkegaard (M1). \cite{mcwhan74,longo70}  
The temperature dependence of the reflections ($\mathrm{\overline{3}}$12), (12$\mathrm{\overline{2}}$), (31$\mathrm{\overline{1}}$), (121) and 
(10$\mathrm{\overline{2}}$)
of the M1 phase that contribute to the diffraction peak at 36.2 1/nm clearly demonstrate that the phase transition was complete. Furthermore, they
demonstrate that our measurements were performed at temperature values at which there is no 
observable phase coexistence due to nanoscale phase separation. \cite{qazilbash11}
From the 2D diffraction images we observed diffuse Debye-Scherrer rings on which brighter spots
were evenly distributed.
We could also observe a small V$_6$O$_{13}$ impurity
component in the diffraction patterns, 
which explains the approximately temperature independent diffraction peaks in the experiment.
The effect of the impurity on the Compton difference profile 
is minor. We note that V$_6$O$_{13}$ undergoes only slight 
thermal expansion between the below- and above-phase-transition measurement temperatures. 
Comparing with previous literature on MgB$_2$, we estimate that this thermal expansion in a pure V$_6$O$_{13}$ sample 
could cause a Compton profile difference of 0.1\% at the maximum, which could result in a 0.001\% Compton profile
difference for 1\% impurity concentration.\cite{nygard04,joshi07}

We measured the Compton scattering spectra repeatedly using 10 minute accumulation times and their mutual agreement within
1$\sigma$ was verified. 
The incident photon flux on each detector element was kept at  $1.5 \times 10^4$ counts per second
using an attenuator coupled to a feedback loop.
The spectra recorded with different analyser 
elements at slightly different scattering angles 
were brought to a common $p_z$ scale with the formula of Holm.\cite{holm88} The shifted spectra 
were then summed up to yield the full spectra 
with $10^7$ counts at the Compton peak.
We applied energy dependent corrections 
for all relevant attenuation sources along the 
incident and scattered beam paths, and for detector efficiency. 
Finally we summed the individual spectra for each detector element. 
The resulting spectra were then normalized to the expected electron number within the
$\pm$ 10 a.u. integration range. We then took the CP differences 
for each analyser, and summed the resulting difference profiles to 
yield the final experimental result. The profile differences we will present from hereon are defined as
$\Delta J(q)=(J(q)_{R}-J(q)_{M1})/J(0)_{R} \times 100\%$.

\section{Computational methods}

The doubly differential scattering cross section for inelastic x-ray scattering is written 
in the Kramers-Heisenberg form as 
\begin{equation}
\frac{d\sigma}{d\Omega d\omega}=(\frac{d\sigma}{d\omega})_{Th}S(\vec{q},\omega),
\end{equation}
where $S(\vec{q},\omega)$ is the dynamical structure factor and Th refers to the Thomson scattering 
cross section. The cross section in the Compton scattering regime, where $\omega$ is much larger than
electron binding energies in the system and 
$\lvert\vec{q}\rvert$ is large compared to the reciprocal lattice unit length scale, can be expressed 
as a projection of the three dimensional EMD onto the scattering vector. This is the impulse 
approximation (IA) as discussed in literature. \cite{eisenberger70,eisenberger74,ribberfors75,holm88,holm89}    
Within the IA the 
scattering cross section is written as 
\begin{equation}
\frac{d\sigma}{d\Omega d\omega}=(\frac{d\sigma}{d\omega})_{Th}\frac{J(p_z)}{q},
\end{equation}
where $J(p_z)=\int dp_xdp_y\rho(p_x,p_y,q=p_z)$.

The Compton profiles for both phases were calculated utilizing density functional theory (DFT) \cite{kohn}, and the full potential augmented plane wave plus local orbitals method, as implemented in the elk code.\cite{elk}
We consider V 3s/3p/3d/4s and O 2s/2p electrons 
as valence electrons that contribute to the momentum densities. 
Exchange and correlation are treated in the local spin density (LSDA and LSDA+U approximations for the R and M1 phases, respectively.
A local Hubbard $U$ correction of $U_{\mathrm{eff}}=U-J$, where U was 
4.2 eV and J was 0.8 eV was applied to the vanadium d-electrons. This selection reproduces the band gap and is widely used in literature. \cite{liebch05,kim13} 
Double counting corrections were done at the fully localised limit. 
The LDA+U ground state of the M1 phase has been studied 
 previously in Ref.~\cite{yuan12}, and we calculated 
the M1 Compton profiles using non-magnetic (NM), 
ferromagnetic (FM) and antiferromagnetic (AFM) solutions of the Kohn-Sham equations, for
later comparison with the experiment. 

We used experimental crystal structures for the R and M1 phases.\cite{longo70,mcwhan74} 
The muffin tin radii were r$_{V,M}$=1.88 a.u., r$_{O,M}$=1.42 a.u., r$_{V,R}$=2.05 a.u.  and r$_{O,R}$=1.53 a.u.  
The interstitial density was represented by plane waves expanded up to R$_{MT_{min}k_{max}}$=8. The 
interstitial densities and potentials were expanded in Fourier components up to $|\mathbf{G}|=22$ atomic units. The 
Kohn-Sham equations were solved on a $8 \times 8 \times 8$ $\Gamma$-centered $\bf{k}$-point grid for both 
phases. The electron momentum densities were calculated and projected onto selected $\vec{q}$ 
directions using a recently presented tetrahedron method. \cite{ernsting14}    
The spherically averaged momentum densities were generated by calculating profiles
for an appropriate distribution of $\vec{q}$ directions. The weighting factors for the averaging 
procedure were generated by a numerical Voronoi division of the Brillouin zones about the 
chosen reciprocal lattice directions. For
the R phase we calculated profiles along non-equivalent
directions with $0 \leq h \leq 5$ and $0 \leq l < k \leq 5$, spanning
the irreducible wedge of the Brillouin zone. For M1 we
chose $ -5 \leq h \leq 5$, $0 \leq k \leq 5$ and $0 \leq l \leq 5$. The
computational profiles were convoluted with a Gaussian
function of 0.7 a.u. full width at half maximum and normalized to 25 electrons, which is the number of valence
electrons per formula unit in the calculation.
We used
the experimental $J_R(0)$ in taking the differences. We
normalize our experimental profiles to the number given
by $\int^{10 a.u.}_{-10 a.u.} dp_z (J_{\textrm{core}}+J_{\textrm{valence}})$, 
where the $J_{\textrm{core}}$ profiles are obtained 
from the Hartree-Fock calculations of Biggs et al.\cite{biggs75}, and the valence contribution 
is fixed to 25 electrons. 

\section{Results and discussion}

\begin{figure}
\includegraphics[width=\linewidth]{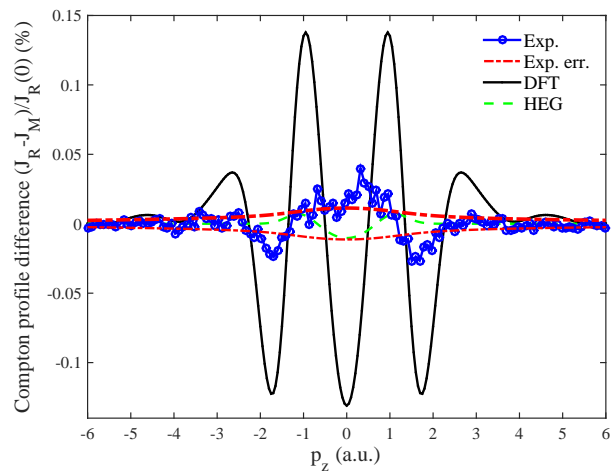}
\caption{\label{fig:expprof}The experimental and 
theoretical differences of the Compton profiles 
of the metallic and insulating phases of poly crystalline VO$_2$. The DFT curve
was calculated using the AFM solution for the M1 phase.}
\end{figure}

The experimental difference profile and the theoretical results are presented in Fig.~\ref{fig:expprof}. 
In the experimental data, one can observe depressions at $p_z = \pm 1.6$ a.~u.~and an 
increase at $p_z=0$. Beyond 4 a.u., the difference is negligible within the experimental 
uncertainty, indicating that only the valence electron states are significantly 
altered across the phase transition, as is expected.
First, we will discuss the difference profile that results 
from considering VO$_2$ as a homogeneous electron
gas with a change in the electron density at the MIT.

For the homogeneous electron gas (HEG), the Compton profile takes a 
simple form: $J(p_z)=\frac{3n_e}{4p_F^3}({p_F}^2-p_z^2)$ for $\lvert{p_z}\rvert \leq p_F$, where ${p_F}$ is 
the Fermi momentum and $n_e$ the number of valence electros per unit cell. The 
density change across the phase transition in the HEG model produces
profile differences via the density dependence of $p_F$. 
The electron densities across the phase transition change by 0.05\%, 
with the rutile phase being denser.
Considering the O 2p and V 3d/4s electrons as the HEG, 
and by using the electron densities based on structural reference data \cite{mcwhan74,longo70}
we obtain the HEG curve in Fig. \ref{fig:expprof}.
We note that the HEG is a poor approximation for the 
valence electrons in oxide materials, but the magnitude 
of the difference in the HEG curve can be considered as an 
indicator of the order of magnitude to expect for the real system. 
We observe that the magnitude of the HEG difference profile is about 
half of the experimental one, and that the sign is incorrect. In first 
approximation, this indicates that the momentum density changes in a way 
that counteracts effects arising from real space densification.

\begin{figure}
\includegraphics[width=\linewidth]{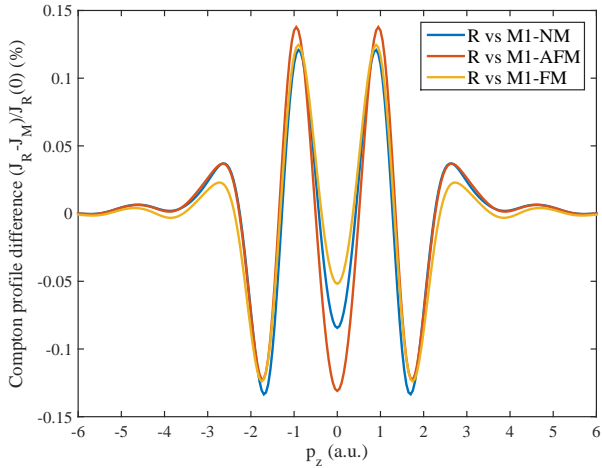}
\caption{\label{fig:compprof}The theoretical Compton profiles differences
for of the metallic and insulating phases of polycrystalline VO$_2$. The DFT curves
were calculated using the NM, FM and AFM solutions for the M1 phase.}
\end{figure}

Second, we will consider the difference profile that we obtain from the LSDA/LSDA+U calculations. 
Taking electronic correlations into account at the LSDA+U level for the M1 phase using the AFM solution,
and LDA level for the R phase, results in a difference profile with an absolute magnitude overestimated 
by a factor of 5 and 
significant changes  beyond 2 a.u.~not observed in the 
experiment. However, the momentum axis positions of the nodes,
minima and maxima in the difference 
profile are in good agreement with experiment. To rule out 
convergence issues as the origin of the observed amplitude-discrepancy, we
calculated individual Compton profiles for a $14 \times 14 \times 14$ k-point meshes for both phases. 
We found that the Compton  difference profiles between R and M1 phases were well converged at the coarser
$8 \times 8 \times 8$ k-point mesh. This is not exactly true for features near $\lvert{p_z}\rvert$ = 0, 
but the convergence is reached in the main features to a satisfactory level, 
and especially well beyond being capable of producing the factor of 5 difference in the amplitude.

There is a number of possible origins of the difference between experiment and theory. 
We consider electron correlation corrections first. 
Previous studies have indicated that errors made in exchange-correlation approximations in DFT tend to 
cancel out in calculations of difference Compton profiles. 
Lehtola et.~al studied the difference profile for a water monomer and a dimer.\cite{lehtola11}
They compared DFT results to highly accurate quantum chemistry 
CCSD calculations, and found a small difference of 0.04\% at $p_z=0$ a.u.
Quantum Monte Carlo calculations of Compton profiles for Si 
and Li have been performed,\cite{louie98,filippi99}
and both studies reached the conclusion that correlation effects are not the primary reason for differences 
between experimental and theoretical Compton profiles.
Recent investigation on dynamical mean field theory corrections to 
Compton profile anisotropies of Ni, for which local correlations are certainly of 
importance due to the unfilled 3d shell of the Ni atom, presented comparisons of LSDA, LSDA+U and 
LSDA+DMFT. Again correlation corrections were observed to induce 
quite small changes in the Compton profile, and the corrections largely cancel out in anisotropy 
profiles. \cite{chioncel14} It it however possible that the error cancellation when comparing two phases
is less complete, as effects of static correlation are more pronounced in the M1 phase. \cite{gatti07,biermann05} 
In fact, after preparing this manuscript for submission, it has come to our knowledge that recent quantum Monte Carlo calculations \cite{kylanpaa18} of the Compton profile difference give much better agreement with the experimental data, indicating that electron-electron interactions play a central role and have to be included more accurately than what is possible within density functional theory and its extensions such as LDA+U. 
Thus we conclude that exchange-correlation effects may play the decisive role in producing the observed discrepancy.

Second, we consider how the calculated difference profile is affected by the presence and
ordering of local moments. The formation of local moments may result in changes in the occupation
numbers of the d-band states and thus can contribute to the discrepancy. 
In Fig.~\ref{fig:compprof} we present a comparison of difference profiles obtained using the
LDA for the R phase, and the LDA+U NM, FM and AFM solutions for the M1 phase. We observe 
that the feature near $p_z=0$ is sensitive to the magnetic nature of the M1 ground state, and that
the features at larger momenta are quite insensitive to it. The R1 phase has been suggested to host fluctuating
local moments. \cite{yuan12} Based on the differences we observe by altering the magnetic properties of the M1 phase, it is possible that the 
R phase profile might be significantly altered 
if the local moments were simulated e.g. using supercell
calculations.

Third, we consider the effects of the unit cell geometries 
on the observed difference between experiment and theory.
Temperature dependent Compton profile experiments
have been performed for the free-electron-like alkali metals and it has 
been observed 
that the CP becomes narrower for increasing temperature, mainly due to the 
reciprocal space shrinking as the real space lattice expands. \cite{sternemann01,huotari02}
A previous study on the temperature dependence of the isotropic Compton profile of MgB$_2$ has 
shown that a unit cell volume expansion of $\approx$ 0.3\% results 
in a CP difference with a similar magnitude as we observed in our experiment.\cite{nygard04,joshi07} 
In the case of LiF it has been observed that approximately 0.7\% change in the unit cell
volume results in an anisotropy Compton profile with maximum amplitude of 0.2\%.\cite{erba15}
Since the density change across the MIT in VO$_2$ equals only to  0.05\%, we conclude that scaling 
the unit cell dimensions by a few per cent would not be able to produce such a dramatic difference.

Lastly, we consider thermal motion. X-ray diffraction has revealed
that the V atoms is VO$_2$ have large root-mean-square thermal displacements in the R-phase. \cite{mcwhan74} 
The effects of phonons on the EMD in sodium has been discussed earlier,\cite{olevano12} 
and the authors argued that it causes momentum density 
redistribution only in the close vicinity of the highest occupied momentum states. 
The large thermal displacements in VO$_2$, that correspond to instantaneous 
deviations from average bond lengths, however suggest that neglect of thermal motion
may be a contributing factor.
It is known from studies of molecular systems that small differences 
in intra-molecular covalent bond lengths can induce large oscillations in 
difference Compton profiles.\cite{hakala06,jcp_nygard07} Nyg{\aa}rd et al.\cite{nygard06} demonstrated 
that an internuclear Li-O distance variation of approximately 10 \% in a Li-H$_2$O cluster could reverse the 
sign of the profile difference at $p_z=0$. A similar study on water also demonstrated that for molecules, 
the profiles are more sensitive to bond lengths than bond angles.\cite{hakala06}
To gain insight on the expected order of magnitude Compton profile difference caused by V-O bond stretching and compression, 
we performed Compton profile calculations on a V$_2$O$_9$H$_{10}$ molecule.  We observed that
modest V-O bond length changes of the order 1\% can cause a CP difference of 
the order  0.1\% at small $\lvert p_z \rvert$. We note that the Compton profile of water-ethanol
mixtures has been shown to be sensitive to bond length differences of the order 0.001 \AA. \cite{juurinen11}
Thus we suggest that the difference may also have a contribution from thermal motion. 

A detailed analysis of the respective roles of the above-mentioned effects possibly explaining 
the observed discrepancy are the subject of future work.
We suggest that the most major contributions 
to the observed discrepancy originate 
from correlation effects beyond LSDA/LSDA+U, and possibly partly owing to thermal motion.
The effects of local moment formation in the R phase may also give significant 
contributions. Thermal motion is especially relevant for the R phase, 
for which large thermal amplitudes of the V atoms have 
been observed. This can lead to overly narrow Compton profile in combination with correlation effects, which 
might explain the large amplitude-difference. For the 
M1 phase the correlation effects are more relevant.  
 
\section{Conclusions}

The effect of the MIT on the electron momentum density of VO$_2$ 
was studied using Compton scattering and DFT calculations. 
We demonstrate that the difference Compton profile  capture a signal of the phase 
transition even in the spherically averaged Compton 
profile. We observed a difference between the 
results of the experiment and theoretical calculations, which we suggest 
to trace back to post-LDA/LDA+U exchange correlation effects and possibly to some extent to thermal motion. The observed discrepancy
highlights the sensitivity of the Compton profile to the approximations
made in first principles simulations.

\section{Authors contributions}
All authors contributed equally to the article.

\section{Acknowledgements}
This work was supported by the Academy of Finland (grants 1283136, 1254065, 1259526, 1295696).
The authors wish to acknowledge CSC --- IT Center for Science, Finland, for computational resources.

\bibliographystyle{epj}
\bibliography{Ruotsalainen_VO2}

\end{document}